\documentclass[aps,showpacs,preprintnumbers,amsmath, amssymb]{revtex4}

\usepackage{float}
\usepackage{graphics,epsfig}
\usepackage{graphicx}
\usepackage{dcolumn}
\usepackage{bm}

\begin{document}
\baselineskip=0.8 cm
\title{{\bf Geometric phases acquired for a two-level atom coupled to fluctuating vacuum scalar fields due to linear acceleration and circular motion }}

\author{Zixu Zhao$^{1}$\footnote{Corresponding author. zhao$_{-}$zixu@yeah.net}, Baoyuan Yang$^{1}$ }
\affiliation{$^{1}$School of Science, Xi'an University of Posts and Telecommunications, Xi'an 710121, China}

\vspace*{0.2cm}
\begin{abstract}
\baselineskip=0.6 cm
\begin{center}
{\bf Abstract}
\end{center}

In open quantum systems, we study the geometric phases acquired for a two-level atom coupled to a bath of fluctuating vacuum massless scalar fields due to linear acceleration and circular motion without and with a boundary. In free space, as we amplify acceleration, the geometric phase acquired purely due to linear acceleration case firstly is smaller than the circular acceleration case in the ultrarelativistic limit for the initial atomic state $\theta\in(0,\frac{\pi}{2})\cup(\frac{\pi}{2},\pi)$, then equals to the circular acceleration case in a certain acceleration, and finally, is larger than the circular acceleration case. The spontaneous transition rates show a similar feature. This result is different from the case of a bath of fluctuating vacuum electromagnetic fields that has been studied. Considering the initial atomic state $\theta \in(0,\pi)$, we find that the geometric phase acquired purely due to linear acceleration always equals to the circular acceleration case for the certain acceleration. The feature implies that, in a certain condition, one can simulate the case of the uniformly accelerated two-level atom by studying the properties of the two-level atom in circular motion. Adding a reflecting boundary, we observe that a larger value of a geometric phase can be obtained compared to the absence of a boundary. Besides, the geometric phase fluctuates along $z$, and the maximum of geometric phase is closer to the boundary for a larger acceleration. We also find that geometric phases can be acquired purely due to the linear acceleration case and circular acceleration case with $\theta \in(0,\pi)$ for a smaller $z$.

\end{abstract}

\maketitle
\newpage
\vspace*{0.2cm}

\section{Introduction}

In quantum mechanics, the Schr$\ddot{o}$dinger equation \cite{Schrodinger} can describe the time evolution of a wave function for a closed system. Considering the special relativity, Klein \cite{Klein} and Gordon \cite{Gordon} obtained a relativistic wave equation (the Klein-Gordon equation), and Dirac \cite{Dirac} gave a wave equation that describes relativistic electrons (the Dirac equation). In 1947, the Lamb shift \cite{Lamb,Bethe} showed that the fine structure of the second quantum state of hydrogen does not agree with the prediction of the Dirac theory and indicated that we need to take into account the interactions between electron and vacuum fluctuations. The Casimir effect \cite{Casimir} revealed that vacuum fluctuations can be modified by the boundaries. From the quantum field theory \cite{BirrellDavies}, the spin-$0$ scalar field and spin-$1/2$ fermion field can be described by the Klein-Gordon equation and Dirac equation respectively. The study of fluctuating vacuum fields has become a very active topic. By treating the atom as an open quantum system in a bath of the fluctuating vacuum electromagnetic fields, the authors showed that the modification of the vacuum fluctuations can be directly detected by means of the measurement of geometric phase \cite{Yu}.

Geometric phase was introduced in Ref. \cite{Pancharatnam}. Aharonov and Susskind \cite{AharonovSusskind} showed the observability of the sign change of spinors under $2\pi$ rotations in certain circumstances. Berry \cite{Berry} found that there is, in addition to the familiar dynamical phase, a geometric phase acquired over the course of a cycle when a quantum system experiences a cyclic adiabatic process. Aharonov and Anandan \cite{AharonovAnandan} introduced a new geometric phase factor that is defined for any cyclic evolution of a quantum system, and this phase factor is a gauge-invariant generalization of the one found by Berry for the special case of adiabatic evolution. Samuel and Bhandari \cite{SamuelBhandari} showed that the geometric phase appears in a more general context, in which the evolution of the quantum system need be neither unitary nor cyclic and may be interrupted by quantum measurements. Every quantum system is an open quantum system because it inevitably interacts with the environment at least with the vacuum fluctuations. In the framework of open quantum systems, the geometric phase for a mixed state under a nonunitary evolution should be considered. Uhlmann \cite{Uhlmann} firstly defined a mixed-state geometric phase through the mathematical concept of purification. Based on the interferometry, Sj$\ddot{o}$qvist $\emph{et}$ $\emph{al}$. \cite{Sjoqvist} gave an alternative definition for the unitarily evolved nondegenerate mixed-state density matrix, and there have been further studies in Refs. \cite{Singh,Tong,WangZS}. The mixed-state geometric phase has been demonstrated in experiments \cite{Du,Ericsson}.

The modification of vacuum fluctuations can be induced by the acceleration of a two-level atom. The Unruh effect \cite{Rindler,Fulling,Hawking1,Hawking2,Davies1,DeWitt,Unruh} showed that, in Minkowski spacetime, a no-particle state of inertial observers corresponds to a thermal state with a temperature $T_{U}=a/(2\pi)$ for uniformly accelerated observers, where $a$ is the observers' proper acceleration, and we adopt natural units $c=\hbar=k_{B}=1$. This is the result of the quantum field theory. Based on the works of Dalibard $\emph{et}$ $\emph{al}$. \cite{DDC1,DDC2}, Audretsch and M$\ddot{u}$ller considered a uniformly accelerated atom coupled to a massless scalar quantum field and calculated the Einstein coefficients for spontaneous excitation and spontaneous emission \cite{AM1} and radiative energy shifts \cite{AM2}. Passante \cite{Passante} showed that the effect of electromagnetic vacuum fluctuations on atomic level shifts is not totally equivalent to that of a thermal field with the temperature $T_{U}=a/(2\pi)$, due to an extra term proportional to $a^2$, contrary to the scalar field case. The Unruh effect has been studied extensively \cite{Benatti2,ZYu,Crispino,Hayden,Adami,Lima,Wang,Tian,Zhao,Zhao2}. However, the effect has not been directly detected. Recently, the authors in Ref. \cite{MFM} showed that the geometric phase can be employed to detect the Unruh effect. Hu and Yu studied the geometric phase for a uniformly accelerated two-level atom coupled in the multipolar scheme to a bath of fluctuating vacuum electromagnetic fields, which provides evidence of the Unruh effect by the phase variation due to the acceleration that can be in principle observed via atom interferometry between the accelerated atom and the inertial one \cite{Hu}. Based on the studies in Refs. \cite{Letaw,Bell,Bell2}, the authors in Ref. \cite{Jin} considered a circularly accelerated two-level atom and found in the ultrarelativistic limit that the phase acquired due to circular motion is always larger than that in linear acceleration with the same proper acceleration for $\theta\in(0,\frac{\pi}{2})\cup(\frac{\pi}{2},\pi)$. In the presence of a boundary, the geometric phase of an accelerated two-level atom coupled to fluctuating vacuum electromagnetic fields has been investigated \cite{Zhai}. There exists a difference between fluctuating vacuum electromagnetic fields and scalar fields. The vacuum of the scalar field in a uniformly accelerated frame is equivalent to a purely thermal field with the temperature $T_{U}=a/(2\pi)$. In order to exclude the extra term proportional to $a^2$, it is therefore very natural to consider a geometric phase acquired for a uniformly accelerated two-level atom coupled to a fluctuating vacuum scalar field. In addition, we would like to know the relation of a geometric phase for a two-level atom between the linear acceleration and circular motion in fluctuating vacuum scalar fields, which can be compared with the result for the electromagnetic field \cite{Jin}. The vacuum fluctuations will be modified because of the presence of the boundary; therefore, it is of great interest to study the geometric phase due to the modification of the vacuum fluctuations caused by the reflecting plane. In this work, we will consider the geometric phases acquired for the atom coupled to a bath of fluctuating vacuum massless scalar fields due to circular motion and linear acceleration without and with a boundary.

The structure of this work is as follows. In Sec. II, we exhibit the basic formula governing dynamical evolution for a two-level atom coupled to the scalar fields in open quantum systems and the equation for the geometric phase. In Sec. III, for a two-level atom coupled to a massless scalar field in the Minkowski vacuum, we compare the geometric phases acquired purely due to linear and circular acceleration without a boundary. In Sec. IV, we study the cases of the presence of a boundary. We will summarize our results in the last section.

\section{The basic formula of dynamical evolution and geometric phase for a two-level atom}

We consider a two-level atom interacting with a bath of fluctuating scalar fields in the Minkowski vacuum. The Hamiltonian of the system is given by
\begin{equation}
H=H_s+H_f+H_I\;,
\end{equation}
where $H_s$, $H_f$, and $H_I$ denote Hamiltonian of the atom, the scalar field, and their interaction. For the atom and the interaction between the atom and the scalar field, their Hamiltonians are given by
\begin{equation}
H_s={\frac{1}{2}}\hbar\omega_0\sigma_3\;,\;\;\; H_I(\tau)=\mu(\sigma_{+}+\sigma_{-})\phi(t,\textbf{x})\;,
\end{equation}
where $\omega_0$ is the energy level spacing of the atom and $\sigma_3$ is the Pauli matrix; $\sigma_{+}$ and $\sigma_{-}$ are the atomic raising and lowering operators, respectively, and $\phi(t,\textbf{x})$ is the operator of the scalar field.

The initial total density matrix of the system takes $\rho_{tot}=\rho(0) \otimes |0\rangle\langle0|$, in which $\rho(0)$ is the initial reduced density matrix of the atom, and $|0\rangle$ is the vacuum state of the field. The evolution of the total density matrix $\rho_{tot}$ reads
\begin{equation}
\frac{\partial\rho_{tot}(\tau)}{\partial\tau}=-\frac{i}{\hbar}[H,\rho_{tot}(\tau)]\;,
\end{equation}
where $\tau$ is the proper time. Assuming that the interaction between the atom and the field is
weak, we obtain the evolution of the reduced
density matrix $\rho(\tau)$ in the Kossakowski-Lindblad form~\cite{Gorini,Lindblad,Benatti1,Benatti3},
\begin{equation}
\frac{\partial\rho(\tau)}{\partial \tau}= -\frac{i}{\hbar}\big[H_{\rm eff},\,
\rho(\tau)\big]
 + {\cal L}[\rho(\tau)]\ ,
\end{equation}
where
\begin{equation}
{\cal L}[\rho]=\frac{1}{2} \sum_{i,j=1}^3
a_{ij}\big[2\,\sigma_j\rho\,\sigma_i-\sigma_i\sigma_j\, \rho
-\rho\,\sigma_i\sigma_j\big]\ .
\end{equation}
We introduce the two-point correlation function for the scalar field,
\begin{equation}
G^{+}(x,x')=\langle0|\phi(t,\textbf{x})\phi(t',\textbf{x}')|0 \rangle\;.
\end{equation}
The Fourier and Hilbert transforms of the field correlation function read, respectively
\begin{equation}
{\cal G}(\lambda)=\int_{-\infty}^{\infty} d\Delta\tau \,
e^{i{\lambda}\Delta\tau}\, G^{+}\big(\Delta\tau\big)\; ,
\quad\quad
{\cal K}(\lambda)=\frac{P}{\pi
i}\int_{-\infty}^{\infty} d\omega\ \frac{ {\cal G}(\omega)
}{\omega-\lambda} \;.
\end{equation}
Therefore, the coefficients of Kossakowski matrix $a_{ij}$ can be written as
\begin{equation}
a_{ij}=A\delta_{ij}-iB
\epsilon_{ijk}\delta_{k3}-A\delta_{i3}\delta_{j3}\;,
\end{equation}
where
\begin{equation}
A=\frac{\mu^2}{4}[{\cal {G}}(\omega_0)+{\cal{G}}(-\omega_0)]\;,\;~~
B=\frac{\mu^2}{4}[{\cal {G}}(\omega_0)-{\cal{G}}(-\omega_0)]\;.\;~~
\end{equation}
By absorbing the Lamb shift term, the effective Hamiltonian $H_{\rm eff}$ is written as
\begin{equation}
H_{\rm eff}=\frac{1}{2}\hbar\Omega\sigma_3=\frac{\hbar}{2}\left\{\omega_0+\frac{i}{2}[{\cal
K}(-\omega_0)-{\cal K}(\omega_0)]\right\}\,\sigma_3\;.
\end{equation}
Under the assumption that the initial state of the atom is
$|\psi(0)\rangle=\cos\frac{\theta}{2}|+\rangle+\sin\frac{\theta}{2}|-\rangle$, one can work out the time-dependent reduced density matrix,

\begin{eqnarray}\label{dens}
\rho(\tau)=\left(
\begin{array}{ccc}
e^{-4A\tau}\cos^2{\frac{\theta}{2}}+{\frac{B-A}{2A}}(e^{-4A\tau}-1) & {\frac{1}{2}}e^{-2A\tau-i\Omega\tau}\sin\theta\\ {\frac{1}{2}}e^{-2A\tau+i\Omega\tau}\sin\theta & 1-e^{-4A\tau}\cos^2{\frac{\theta}{2}}-{\frac{B-A}{2A}}(e^{-4A\tau}-1)
\end{array}\right)\;.
\end{eqnarray}
The geometric phase for a mixed state under a nonunitary evolution
can be defined as~\cite{Tong}
\begin{equation}
 \label{gp} \gamma_{g}=\arg \left(
\sum\limits_{k=1}^N \sqrt{\lambda_k(0)\lambda_k(T)}\langle
\phi_k(0)|\phi_k(T)\rangle e^{-\int_0^T \langle \phi_k(\tau)|\dot
\phi_k(\tau) \rangle d\tau} \right)\;,
\end{equation}
where $\lambda_k(\tau)$ and $|\phi_k(\tau)\rangle$ are the eigenvalues and eigenvectors of
the reduced density matrix $\rho(\tau)$. In order to find the
geometric phase,  we first calculate the eigenvalues of the density
matrix (\ref{dens}) to get $
\lambda_\pm(\tau)={\frac{1}{2}}(1\pm\eta)\;,$ where
$\eta=\sqrt{\rho_3^2+e^{-4A\tau}\sin^2\theta}$ and
$\rho_3=e^{-4A\tau}\cos\theta+{B\over A}(e^{-4A\tau}-1)$. It is easy to see
that $\lambda_-(0)=0$. Therefore, the contribution only
comes from the eigenvector corresponding to $\lambda_+$ ,
\begin{equation}
|\phi_+(\tau)\rangle=\sin{\frac{\theta_{\tau}}{2}}|+\rangle+\cos{\frac{\theta_{\tau}}{2}}e^{i\Omega\tau}|-\rangle\;,
\end{equation}
where
\begin{equation}
\tan{\frac{\theta_{\tau}}{2}}=\sqrt{\frac{\eta+\rho_3}{\eta-\rho_3}}\;.
\end{equation}
The geometric phase can then be calculated directly using Eq.~(\ref{gp})
\begin{equation}
\gamma_{g}=-\Omega\int_0^T\cos^2{\frac{\theta_{\tau}}{2}}\,d\tau\;.
\end{equation}
Therefore, the geometric phase can be written as
\begin{equation}
\gamma_{g}=-\int_0^{T}{1\over2}\bigg(1-\frac{R-Re^{4A\tau}+\cos\theta}{\sqrt{e^{4A\tau}\sin^2\theta
+(R-Re^{4A\tau}+\cos\theta)^2}}\bigg)\,\Omega\,d\tau\;,
\end{equation}
where $R=B/A$. For a single period of evolution, the  result of this integral can be expressed as
\begin{equation}
\gamma_{g}={\frac{\Omega}{\omega_0}}[F(2\pi)-F(0)]\;,
\end{equation}
where the function $F(\varphi)$ is defined as
\begin{eqnarray}
F(\varphi)&=&-{\frac{1}{2}}\,\varphi-{\frac{\omega_0}{8A}}\ln \left({{\frac{1-Q^2-R^2+2R^2e^{4A\varphi/\omega_0}}{2R}}}+S(\varphi) \right)\nonumber\\
&&-{\frac{\omega_0}{8A}}\text{sgn}(Q)\ln \left({ {1-Q^2-R^2+2Q^2e^{-4A\varphi/\omega_0} }+2|Q|\,S(\varphi)\,{e^{-4A\varphi/\omega_0}}} \right)\;,
\end{eqnarray}
where $S(\varphi)=\sqrt {R^2 e^{8A\varphi/\omega_0}+(1-Q^2-R^2){e^{4A\varphi/\omega_0}}+Q^2}\;$, $Q=R+\cos\theta$, and $\text{sgn}(Q)$ is the standard sign function. We adopt natural units $c=\hbar=k_{B}=1$.

\section{Geometric phases acquired for a two-level atom due to linear acceleration and circular motion in free space}

For a uniformly accelerated two-level atom coupled to a massless scalar field in the Minkowski vacuum, the trajectory of the atom can be described as
\begin{eqnarray}
t{(\tau)}={\frac{1}{a}}\sinh a\tau\;,~~~
x{(\tau)}={\frac{1}{a}}\cosh a\tau\;,~~~
y{(\tau)}=y_{0}\;,~~~
z{(\tau)}=z_{0}\label{trajl}\;.
\end{eqnarray}

The Wightman function of massless scalar field in the Minkowski vacuum takes the form,

\begin{eqnarray}
G^{+}_{0}(x,x')=-\frac{1}{4\pi^2}
           \frac{1}{(t-t'-i\varepsilon)^2-(x-x')^2-(y-y')^2-(z-z')^2}\;.
\end{eqnarray}
Applying the trajectory of the atom (\ref{trajl}), we obtain the field correlation function
\begin{eqnarray}
G^{+}_{0l}(x,x')=-\frac{a^2}{16\pi^2}\frac{1}{\sinh^2(\frac{a\Delta\tau}{2}-i\varepsilon)}\;,
\end{eqnarray}
where $\Delta\tau=\tau-\tau'$.

The Fourier transform of the field correlation function is given by
\begin{eqnarray}
{\cal G}_{0l}(\lambda)=\frac{1}{2\pi}\frac{\lambda}{1-e^{-2\pi\lambda/a}}\;.
\end{eqnarray}
The coefficients for the Kossakowski matrix are
\begin{eqnarray}
\begin{aligned}
&A_{0l}={\gamma_0\over4}\,\coth\frac{\pi\omega_0}{a}\;,\\
&B_{0l}={\gamma_0\over4}\;,\\
\end{aligned}
\end{eqnarray}
where $\gamma_0=\mu^2\omega_0/(2\pi)$ is the spontaneous emission rate.
We obtain the geometric phase,
\begin{eqnarray}
\gamma_{gl}\approx-\pi(1-\cos \theta)-\pi^2\frac{\gamma_0}{2\omega_0}\sin^2\theta \left(2+\cos\theta+\frac{2}{e^{\frac{2\pi \omega_0}{a}}-1}\cos \theta\right).
\end{eqnarray}
The first term $-\pi(1-\cos \theta)$ is the geometric phase for a closed system under unitary evolution. The second term is a correction due to the interaction between the linearly accelerated atom and the environment. For the fluctuating vacuum electromagnetic field in the multipolar scheme, the geometric phase contains the term $\sim (1+{a^2}/{c^2\omega_{0}^2})$ , which has an extra term proportional to $a^2$ \cite{Hu}. The similar $a^2$ term also appeared in Ref. \cite{Passante}, and Passante showed that the presence of this term is a direct consequence of the $\sim a^4 \sinh^{-4}[a(\tau-\tau')/2c]$ behavior of the symmetric correlation function, which is different from the term $\sim a^2 \sinh^{-2}[a(\tau-\tau')/2c]$ for the scalar field case.

For the limit of $a\rightarrow0$, we get the inertial atom case,
\begin{eqnarray}
\gamma_{gi}\approx-\pi(1-\cos \theta)-\pi^2\frac{\gamma_0}{2\omega_0}\sin^2\theta \left(2+\cos \theta\right),
\end{eqnarray}
in which there also exists a correction because of the zero point fluctuations of the Minkowski vacuum. The correction to the geometric phase purely due to the linearly accelerated motion can be obtained
\begin{eqnarray}\label{gpl}
\delta_{0l} =\gamma_{gl}-\gamma_{gi}\approx-\pi^2\frac{\gamma_0}{2\omega_0} \frac{2}{e^{\frac{2\pi \omega_0}{a}}-1}\sin^2\theta \cos \theta.
\end{eqnarray}

Now we calculate the geometric phase of a two-level atom with the circular motion. The trajectory of the atom can be expressed as
\begin{eqnarray}
t(\tau)=\gamma\tau\ ,\ \ \ x(\tau)=r\cos \frac{\gamma\tau v}{r}\ ,\ \
\ y(\tau)=r\sin \frac{\gamma\tau v}{r}\ ,\ \ \ z(\tau)=0\label{trajc}\;,
\end{eqnarray}
where $v$ denotes the tangential velocity of the circularly accelerated atom, $r$ is the radius of the orbit, and $\gamma={1}/{\sqrt{1-{v^2}}}$ is the Lorentz factor. The centripetal acceleration of the atom is $a={\gamma^2v^2}/{r}$. Applying the trajectory of the atom (\ref{trajc}), we need expand $\sin^2[a\Delta\tau/(2v{\gamma})]=\frac{a^2{(\Delta\tau)}^2}{4v^2{\gamma}^2}
-\frac{a^4{(\Delta\tau)}^4}{48v^4{\gamma}^4}+\frac{a^6{(\Delta\tau)}^6}{1440v^6{\gamma}^6}-...$ with $\Delta\tau=\tau-\tau'$. Since it is hard to find the explicit form of ${\cal G}(\omega_0)$ and ${\cal G}(-\omega_0)$ with all orders of $\Delta\tau$, we consider the ultrarelativistic limit $\gamma\gg1$~\cite{Bell}, in which
\begin{eqnarray}
G^{+}_{0c}(x,x')=-\frac{1}{4\pi^2} \frac{1}{(\Delta\tau-i\varepsilon)^2[1+a^{2}(\Delta\tau-i\varepsilon)^2/12]}\;.
\end{eqnarray}
Therefore, the Fourier transforms of the field correlation function are
\begin{eqnarray}
{\cal G}_{0c}(\omega_0)=\frac{\mu^2\omega_0}{2\pi}\bigg(1+\frac{a }{4\sqrt{3}\omega_0}e^{-{\frac{2\sqrt{3}\omega_0}{a}}}\bigg)\;,
\end{eqnarray}

\begin{eqnarray}
{\cal G}_{0c}(-\omega_0)=\frac{\mu^2\omega_0}{2\pi}\frac{a }{4\sqrt{3}\omega_0}e^{-{\frac{2\sqrt{3}\omega_0}{a}}}\;.
\end{eqnarray}
We can obtain
\begin{eqnarray}
\begin{aligned}
&A_{0c}={\gamma_0\over4}\bigg(1+\frac{a }{2\sqrt{3}\omega_0}e^{-{\frac{2\sqrt{3}\omega_0}{a}}}\bigg)\;,\\
&B_{0c}={\gamma_0\over4}.
\end{aligned}
\end{eqnarray}

\begin{figure}[ht]
\includegraphics[scale=0.55]{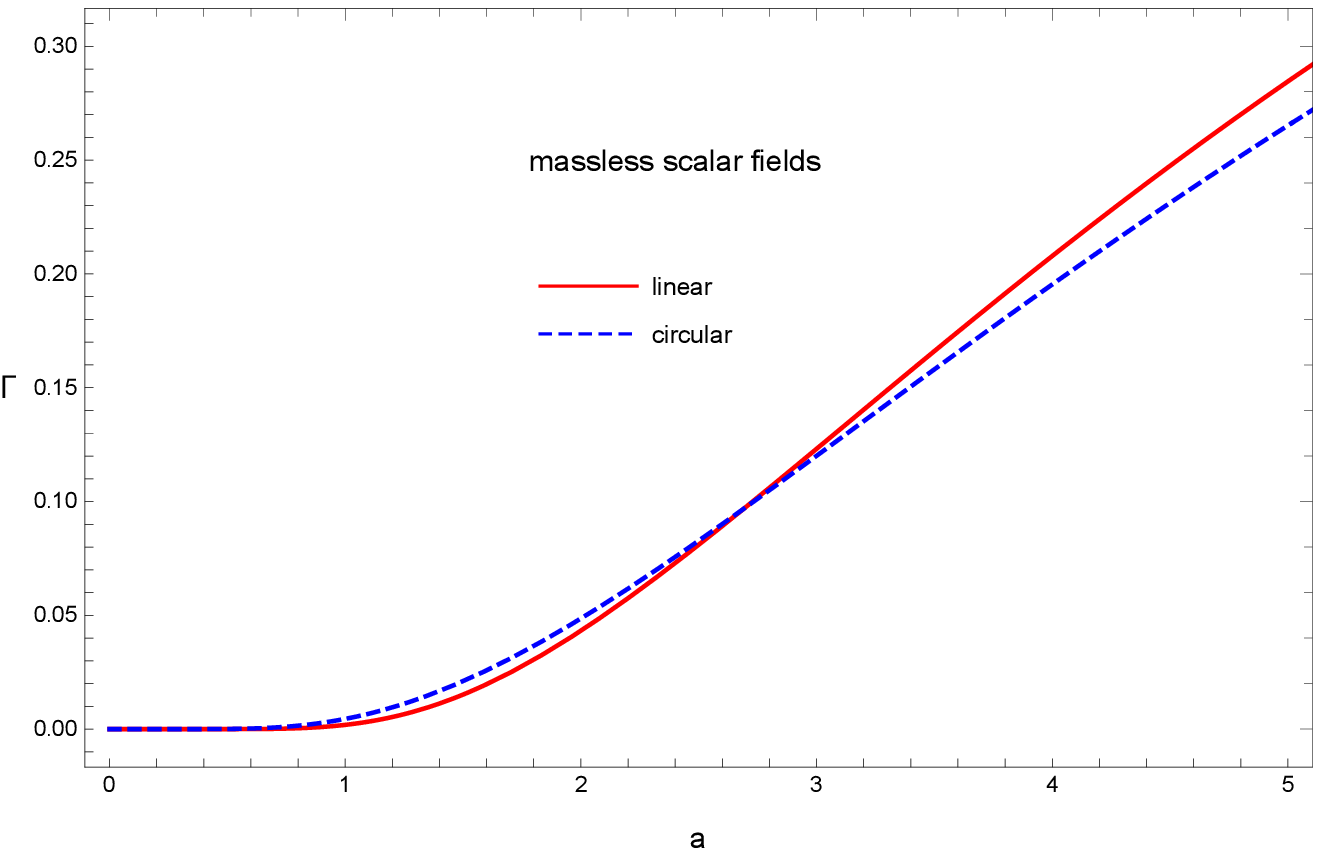}\vspace{0.0cm}
\includegraphics[scale=0.55]{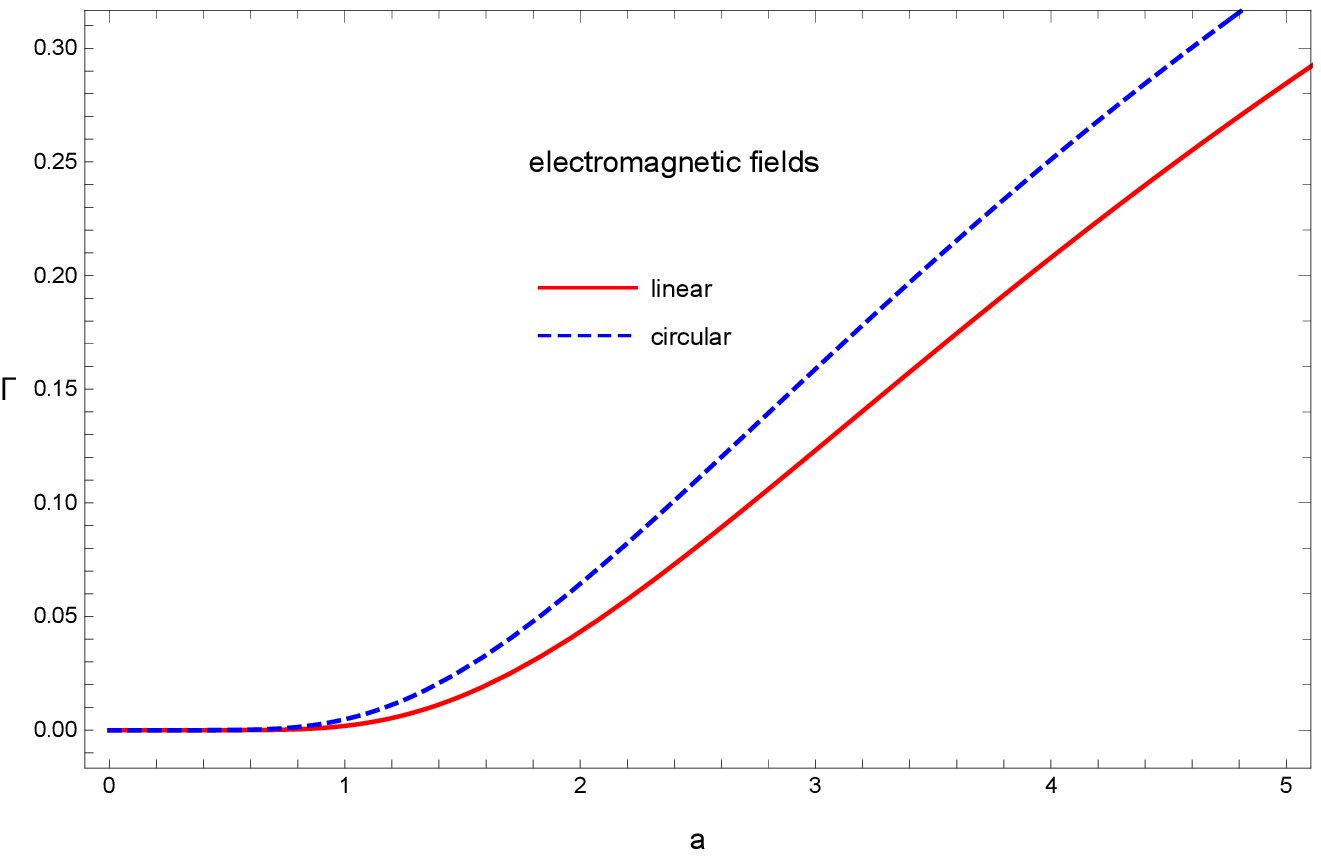}\vspace{0.0cm}
\caption{\label{Rtr} Relative transition rate as a function of acceleration. The left panel and the right panel correspond the vacuum massless scalar fields and electromagnetic fields. The solid (red) and dashed (blue) lines represent the linearly and circularly accelerated atom respectively.}
\end{figure}

In the left panel of Fig. \ref{Rtr}, we present the relative transition rate $\Gamma={\cal G}_{0}(-\omega_0)/{\cal G}_{0}(\omega_0)$ as a function of the acceleration of the atom in vacuum massless scalar fields. We find that the relative transition rate in the linear acceleration case firstly is smaller than circular acceleration case, then equals to the circular acceleration case in a certain $a$, and finally, is larger than the circular acceleration case. The result is different from the case of fluctuating vacuum electromagnetic fields that has been studied in Ref. \cite{Jin}, where the relative transition rate in the circular acceleration case is always larger than that in the linear acceleration case, as shown graphically in the right panel of Fig. \ref{Rtr}.

The geometric phase for the circular motion can be written as
\begin{eqnarray}
\gamma_{gc}\approx-\pi(1-\cos \theta)-\pi^2\frac{\gamma_0}{2\omega_0}\sin^2\theta \left(2+\cos\theta+\frac{a}{2\sqrt{3}\omega_0}\cos \theta e^{-{\frac{2\sqrt{3}\omega_0}{a}}}\right).
\end{eqnarray}
For the limit of $a\rightarrow0$, we get the inertial atom case $\gamma_{gi}$. The correction to the geometric phase purely due to the circular motion can be found by subtracting the contribution of the inertial part $\gamma_{gi}$,
\begin{eqnarray}\label{gpc}
\delta_{0c} =\gamma_{gc}-\gamma_{gi}\approx-\pi^2\frac{\gamma_0}{2\omega_0} \frac{a}{2\sqrt{3}\omega_0} e^{-{\frac{2\sqrt{3}\omega_0}{a}}}\sin^2 \theta\cos \theta.
\end{eqnarray}
It should be noted that there exists a certain acceleration $a$, where the linear and circular accelerations lead to the same geometric phase acquired for every $\theta$. The certain $a$ can be calculated by using Eqs. (\ref{gpl}) and (\ref{gpc}),
\begin{eqnarray}
\frac{2}{e^{\frac{2\pi \omega_0}{a}}-1}=\frac{a}{2\sqrt{3}\omega_0} e^{-{\frac{2\sqrt{3}\omega_0}{a}}}.
\end{eqnarray}

In the following discussion, we use $a\rightarrow \tilde{a}\equiv{a}/{\omega_0}$,~$\delta_{0l} \rightarrow \tilde{\delta_{0l}}\equiv\delta_{0l}/(\frac{\pi^2\gamma_0}{2\omega_0})$,~$\delta_{0c} \rightarrow \tilde{\delta_{0c}}\equiv\delta_{0c}/(\frac{\pi^2\gamma_0}{2\omega_0})$. For simplicity, $\tilde{a}$, $\tilde{\delta_{0l}}$ and $\tilde{\delta_{0c}}$ will be written as $a$, ${\delta_{0l}}$ and ${\delta_{0c}}$.

\begin{figure}[ht]
\includegraphics[scale=0.55]{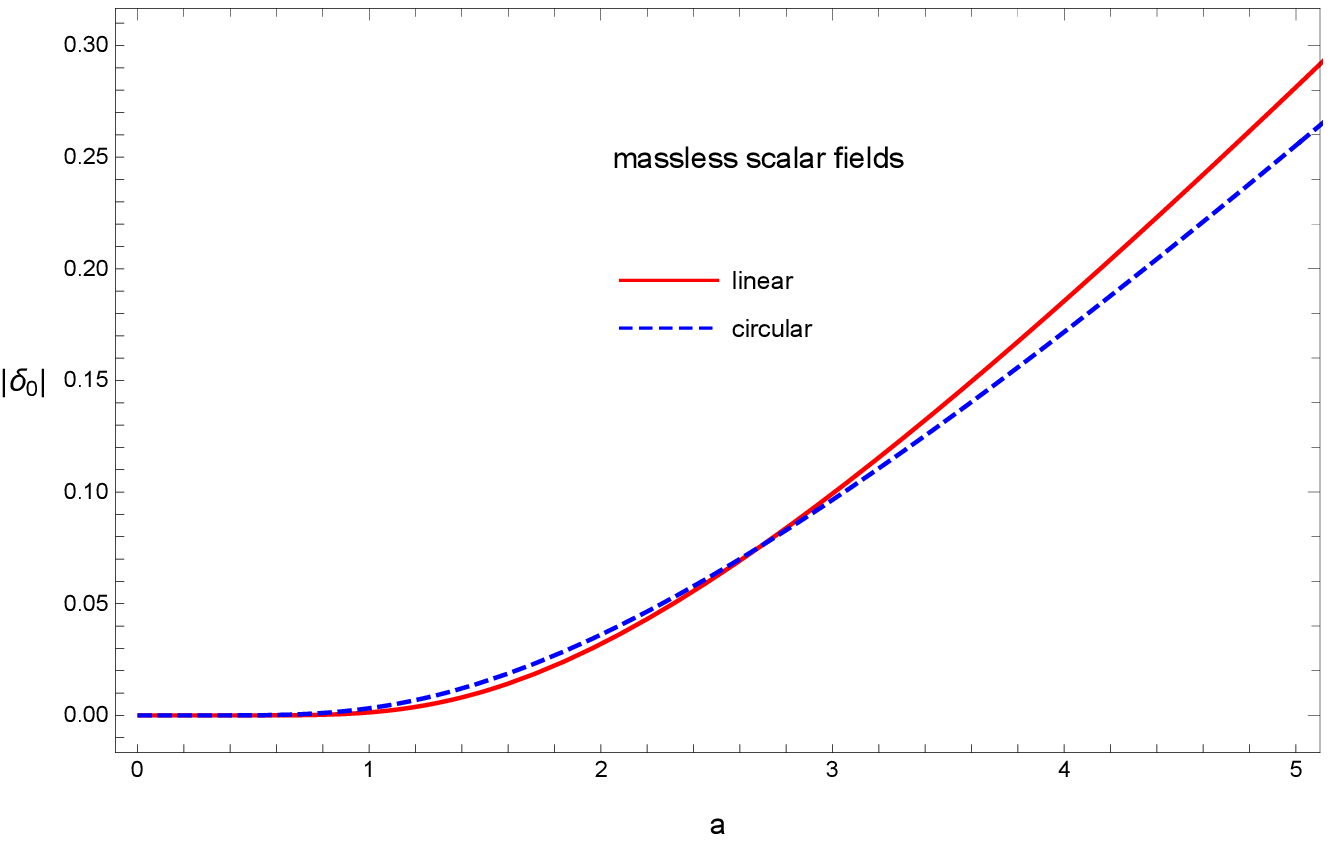}\vspace{0.0cm}
\includegraphics[scale=0.55]{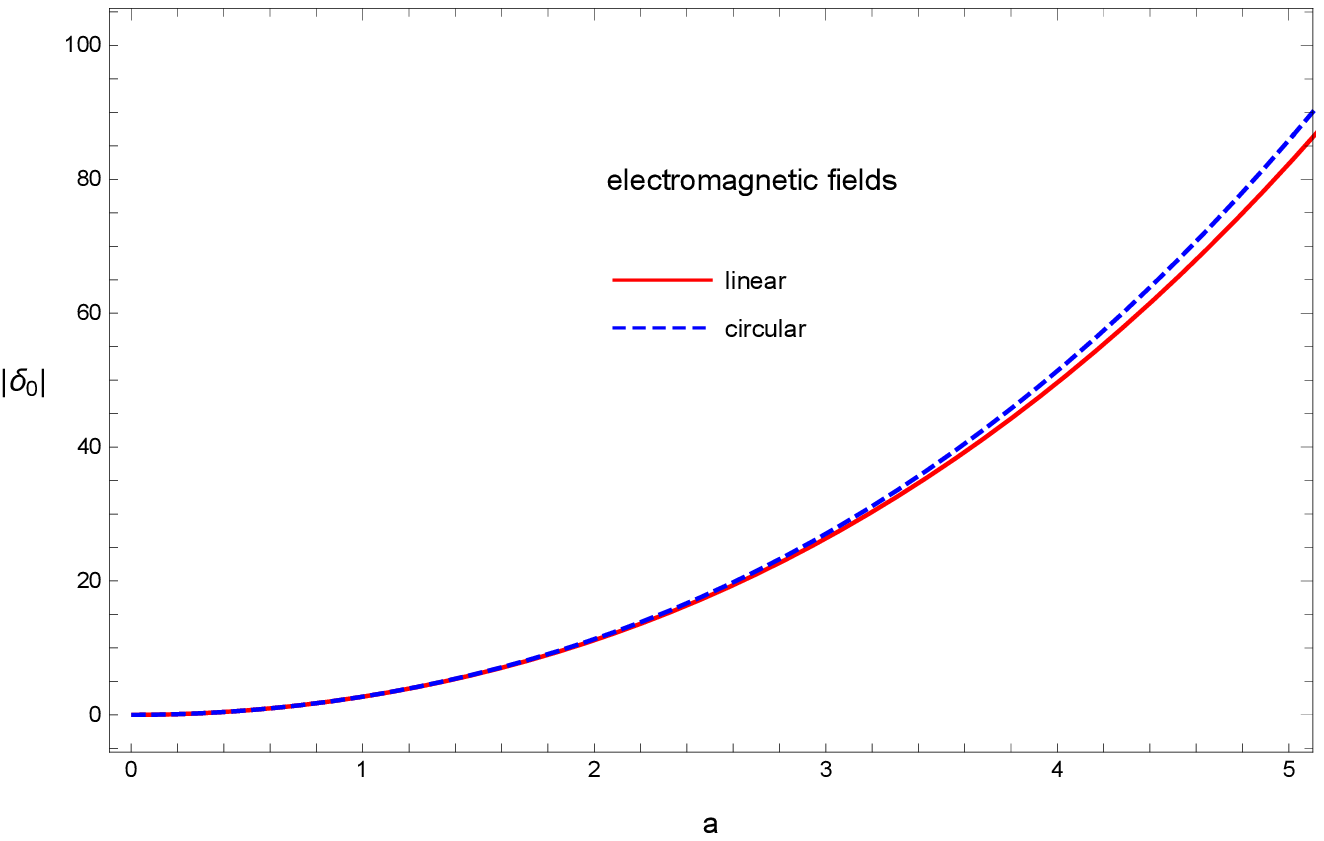}\vspace{0.0cm}
\caption{\label{gpa} Geometric phase as a function of acceleration for the initial atomic state $\theta=\pi/4$. The left panel and the right panel correspond the vacuum massless scalar fields and electromagnetic fields. The solid (red) and dashed (blue) lines represent the geometric phases purely due to linear acceleration case and circular acceleration case respectively.}
\end{figure}

In the left panel of Fig. \ref{gpa}, we describe the geometric phase as a function of acceleration with the initial atomic state $\theta=\pi/4$ for the vacuum massless scalar fields. Increasing the acceleration $a$, we find that the geometric phase acquired purely due to linear acceleration firstly is smaller than circular acceleration case, then equals to the circular acceleration case in a certain $a$, and finally, is larger than the circular acceleration case. This result is different from the case of fluctuating vacuum electromagnetic fields that has been studied in Ref. \cite{Jin}, where the phase acquired purely due to circular acceleration case is always larger than that due to linear acceleration, as shown in the right panel of Fig. \ref{gpa}. We deduce that the relation between linear acceleration case and circular acceleration case is not the same for different fluctuating vacuum fields.

\begin{figure}[ht]
\includegraphics[scale=0.57]{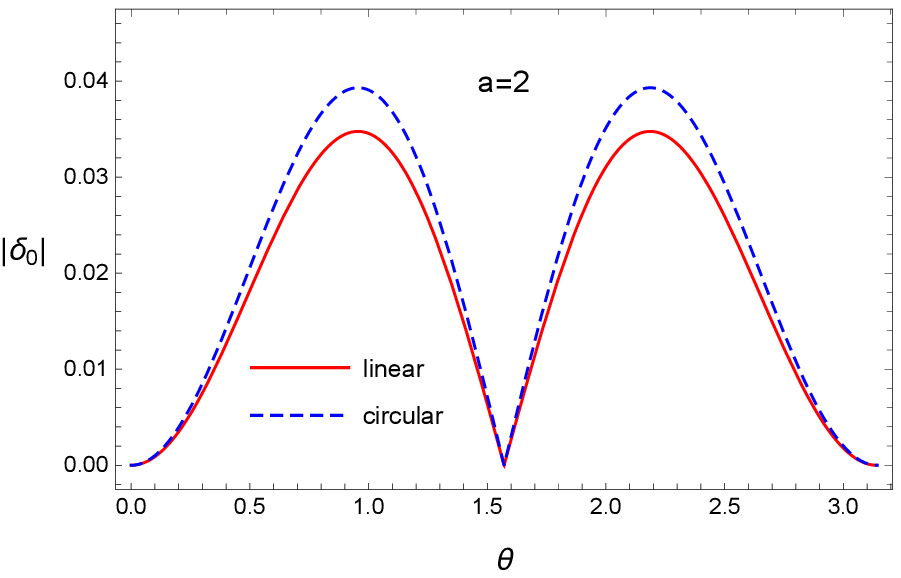} \vspace{0.0cm}
\includegraphics[scale=0.57]{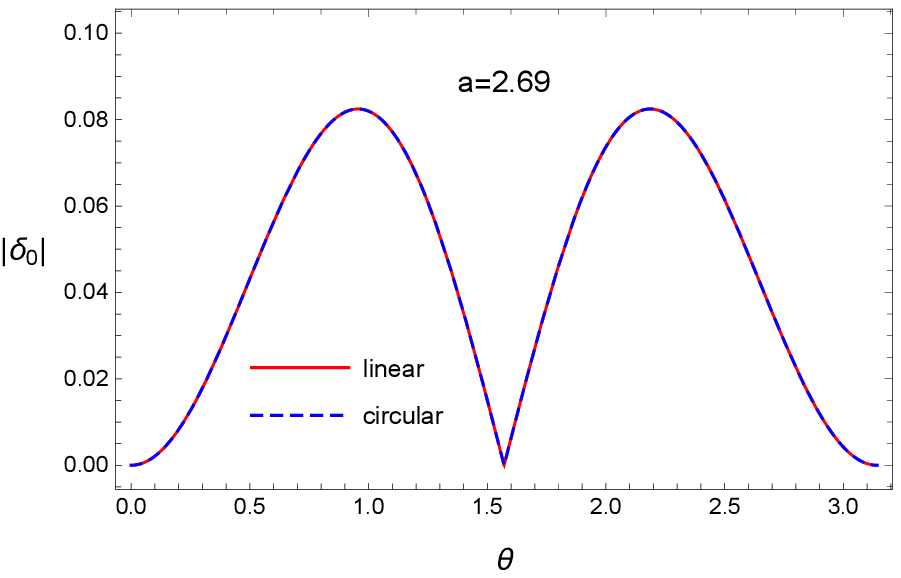}\vspace{0.0cm}
\includegraphics[scale=0.57]{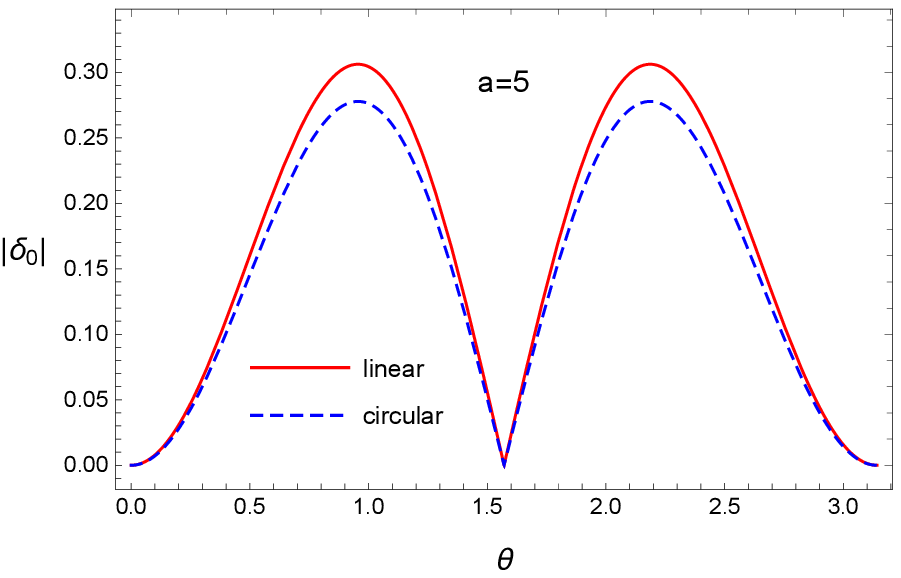}\\ \vspace{0.0cm}
\caption{\label{gptheta} Geometric phase as a function of the initial atomic state for $a=2$, $a=2.69$ and $a=5$. The solid (red) and dashed (blue) lines represent the geometric phases purely due to linear acceleration case and circular acceleration case respectively.}
\end{figure}

In Fig. \ref{gptheta}, we plot the geometric phase as a function of the initial atomic state for different acceleration $a$. We observe that the geometric phases are the periodic function of the initial atomic state $\theta$. We find that the geometric phase acquired purely due to linear acceleration equals to the circular acceleration case when $a\approx2.69$ for every initial atomic state $\theta \in(0,\pi)$. The very large acceleration required for experiments is more feasible to achieve in the circular motion. For the massless scalar field, by taking the particular acceleration $a$, one can use circularly accelerated atom to simulate linearly accelerated atom where the Unruh temperature is $T_{U}=a/(2\pi)$.

The vacuum fluctuations will be modified because of the presence of a boundary. Therefore, we would like to know the geometric phase of an atom in the vicinity of a reflecting boundary.

\section{Geometric phases acquired for a two-level atom due to linear acceleration and circular motion with a boundary}

We add a boundary at $z=0$ and consider a uniformly accelerated atom moving in the $x-y$ plane at a distance $z$ from the boundary. Then, the two-point function in this case can be expressed as
\begin{eqnarray}
G^+(x,x')
=G^{+}_{0}(x,x')+G^{+}_{b}(x,x')\;,
\end{eqnarray}
where $G^{+}_{0}(x,x')$ is the two-point function in the unbounded case, which has already been calculated above, and
\begin{eqnarray}
G^{+}_{b}(x,x')&=&-\frac{1}{4\pi^2}
\frac{1}{(x-x')^2+(y-y')^2+(z+z')^2-(t-t'-i\varepsilon)^2},\;
\end{eqnarray}
gives the correction due to the presence of the boundary.
Applying the trajectory of the atom (\ref{trajl}), we obtain the field correlation function,
\begin{eqnarray}
G^{+}_{l}(x,x')=-\frac{a^2}{16\pi^2}\left[\frac{1}{\sinh^2(\frac{a\Delta\tau}{2}-i\varepsilon)}-\frac{1}{\sinh^2(\frac{a\Delta\tau}{2}-i\varepsilon)-a^2 z^2}\right],
\end{eqnarray}
where $\Delta\tau=\tau-\tau'$.
The Fourier transform of the field correlation function is given by
\begin{eqnarray}
{\cal G}_{l}(\lambda)=\frac{1}{2\pi}\frac{\lambda}{1-e^{-2\pi\lambda/a}}-\frac{1}{2\pi}\frac{\lambda}{1-e^{-2\pi\lambda/a}}\frac{\sin[\frac{2\lambda}{a}\sinh^{-1}(az)]}{2z\lambda\sqrt{1+a^2 z^2}}.
\end{eqnarray}
The coefficients for the Kossakowski matrix are
\begin{eqnarray}
\begin{aligned}
&A_{l}={\gamma_0\over4}\coth\frac{\pi\omega_0}{a}\left\{1-\frac{\sin[\frac{2\omega_0}{a}\sinh^{-1}(az)]}{2z\omega_0\sqrt{1+a^2 z^2}}\right\}     \;,\\
&B_{l}={\gamma_0\over4}\left\{1-\frac{\sin[\frac{2\omega_0}{a}\sinh^{-1}(az)]}{2z\omega_0\sqrt{1+a^2 z^2}}\right\} \;.\\
\end{aligned}
\end{eqnarray}
Then, the geometric phase can be written as
\begin{eqnarray}
\gamma_{bl}\approx-\pi(1-\cos \theta)-\pi^2\frac{\gamma_0}{2\omega_0}\sin^2\theta (2+\cos\theta \coth\frac{\pi \omega_0}{a})\left\{1-\frac{\sin[\frac{2\omega_0}{a}\sinh^{-1}(az)]}{2z\omega_0\sqrt{1+a^2z^2}}\right\}.
\end{eqnarray}
For the limit of $a\rightarrow0$, we obtain the inertial atom case,
\begin{eqnarray}
\gamma_{bi}\approx-\pi(1-\cos \theta)-\pi^2\frac{\gamma_0}{2\omega_0}\sin^2\theta (2+\cos\theta)\left[1-\frac{\sin (2z\omega_0)}{2z\omega_0}\right].
\end{eqnarray}
The correction to the geometric phase purely due to the linearly accelerated motion can be written as
\begin{eqnarray}
\delta_{bl}\approx-\pi^2\frac{\gamma_0}{2\omega_0}\sin^2\theta\left\{ (2+\cos\theta \coth\frac{\pi \omega_0}{a})\left[1-\frac{\sin[\frac{2\omega_0}{a}\sinh^{-1}(az)]}{2z\omega_0\sqrt{1+a^2z^2}}\right]-(2+\cos\theta)\left[1-\frac{\sin (2z\omega_0)}{2z\omega_0}\right] \right\}.
\end{eqnarray}

Now we consider a two-level atom with the circular motion moving in the $x-y$ plane at a distance $z$ from the boundary. We have
\begin{eqnarray}
G^{+}_{bc}(x,x')=-\frac{1}{4\pi^2} \frac{1}{4z^2-(\Delta\tau-i\varepsilon)^2-a^2(\Delta\tau-i\varepsilon)^4/12}\;.
\end{eqnarray}
The Fourier transforms of the correlation function can be written as
\begin{eqnarray}
{\cal G}_{c}(\omega_0)&=&\frac{\omega_0}{2\pi}\bigg[1+\frac{a }{4\sqrt{3}\omega_0}e^{-{\frac{2\sqrt{3}\omega_0}{a}}}-\frac{\sqrt{3}a}{\sqrt{-3+\sqrt{9+12a^2z^2}}{\sqrt{6+8a^2z^2}\omega_0}}\sin\frac{\sqrt{-6+2\sqrt{9+12a^2z^2}}\omega_0}{a}\nonumber\\
&-&\frac{\sqrt{3}a}{2\sqrt
{3+\sqrt{9+12a^2z^2}}{\sqrt
{6+8a^2z^2}\omega_0}}e^{-\frac{\sqrt
{6+2\sqrt{9+12a^2z^2}}\omega_0}{a}}\bigg]\;,
\end{eqnarray}

\begin{equation}
{\cal G}_{c}(-\omega_0)=\frac{\omega_0}{2\pi}\bigg[\frac{a }{4\sqrt{3}\omega_0}e^{-{\frac{2\sqrt{3}\omega_0}{a}}}-\frac{\sqrt{3}a}{2\sqrt
{3+\sqrt{9+12a^2z^2}}{\sqrt
{6+8a^2z^2}\omega_0}}e^{-\frac{\sqrt
{6+2\sqrt{9+12a^2z^2}}\omega_0}{a}}\bigg]\;.
\end{equation}
We therefore have
\begin{eqnarray}
A_{c}&=&{\gamma_0\over4}\bigg[1+\frac{a }{2\sqrt{3}\omega_0}e^{-{\frac{2\sqrt{3}\omega_0}{a}}}-\frac{\sqrt{3}a}{\sqrt{-3+\sqrt{9+12a^2z^2}}{\sqrt{6+8a^2z^2}\omega_0}}\sin\frac{\sqrt{-6+2\sqrt{9+12a^2z^2}}\omega_0}{a}\nonumber\\
&-&\frac{\sqrt{3}a}{\sqrt
{3+\sqrt{9+12a^2z^2}}{\sqrt
{6+8a^2z^2}\omega_0}}e^{-\frac{\sqrt
{6+2\sqrt{9+12a^2z^2}}\omega_0}{a}}\bigg]\;,\nonumber\\
B_{c}&=&{\gamma_0\over4}\bigg[1-\frac{\sqrt{3}a}{\sqrt{-3+\sqrt{9+12a^2z^2}}{\sqrt{6+8a^2z^2}\omega_0}}\sin\frac{\sqrt{-6+2\sqrt{9+12a^2z^2}}\omega_0}{a}\bigg].
\end{eqnarray}
We obtain the geometric phase,
\begin{eqnarray}
\gamma_{bc}&\approx&-\pi(1-\cos \theta)-\pi^2\frac{\gamma_0}{2\omega_0}\sin^2\theta \bigg[2+\cos\theta+\frac{a}{2\sqrt{3}\omega_0}\cos \theta e^{-{\frac{2\sqrt{3}\omega_0}{a}}}\nonumber\\
&-&\frac{\sqrt{6}a}{2V\omega_0}(\frac{\cos\theta}{\sqrt{U+3}} e^{-\frac{\omega_0\sqrt{2U+6}}{a}}+\frac{(2+\cos\theta)}{\sqrt{U-3}}\sin\frac{\omega_0\sqrt{2U-6}}{a})\bigg]\;,\nonumber\\
\end{eqnarray}
where
\begin{eqnarray}
U=\sqrt{9+12a^2z^2},~V=\sqrt{3+4a^2z^2}.
\end{eqnarray}
We get the inertial atom case $\gamma_{bi}$ for $a\rightarrow0$. The correction to the geometric phase purely due to the circular motion can be obtained
\begin{eqnarray}
\delta_{bc}& =&\gamma_{bc}-\gamma_{bi}\approx-\pi^2\frac{\gamma_0}{2\omega_0}\sin^2\theta \bigg\{\frac{a}{2\sqrt{3}\omega_0}\cos \theta e^{-{\frac{2\sqrt{3}\omega_0}{a}}}\nonumber\\
&-& \frac{\sqrt{6}a}{2V\omega_0}\left[\frac{\cos\theta}{\sqrt{U+3}} e^{-\frac{\omega_0\sqrt{2U+6}}{a}}+\frac{(2+\cos\theta)}{\sqrt{U-3}}\sin\frac{\omega_0\sqrt{2U-6}}{a}\right]+(2+\cos\theta)\frac{\sin (2z\omega_0)}{2z\omega_0}\bigg\}\;.
\end{eqnarray}

In the following discussion, we use $a\rightarrow \tilde{a}\equiv{a}/{\omega_0}$,~$\delta_{bl} \rightarrow \tilde{\delta_{bl}}\equiv\delta_{bl}/(\frac{\pi^2\gamma_0}{2\omega_0})$,~$\delta_{bc} \rightarrow \tilde{\delta_{bc}}\equiv\delta_{bc}/(\frac{\pi^2\gamma_0}{2\omega_0})$,~$z\rightarrow \tilde{z}\equiv z \omega_0$. For simplicity, $\tilde{a}$, $\tilde{\delta_{bl}}$,$\tilde{\delta_{bc}}$ and $\tilde{z}$ will be written as $a$, ${\delta_{bl}}$, ${\delta_{bc}}$, and $z$.

\begin{figure}[ht]
\includegraphics[scale=0.57]{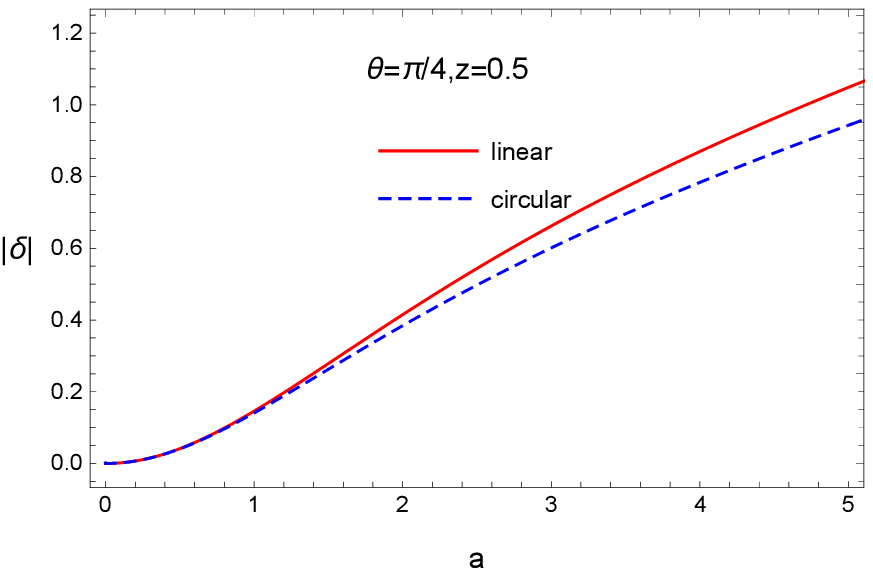} \vspace{0.0cm}
\includegraphics[scale=0.57]{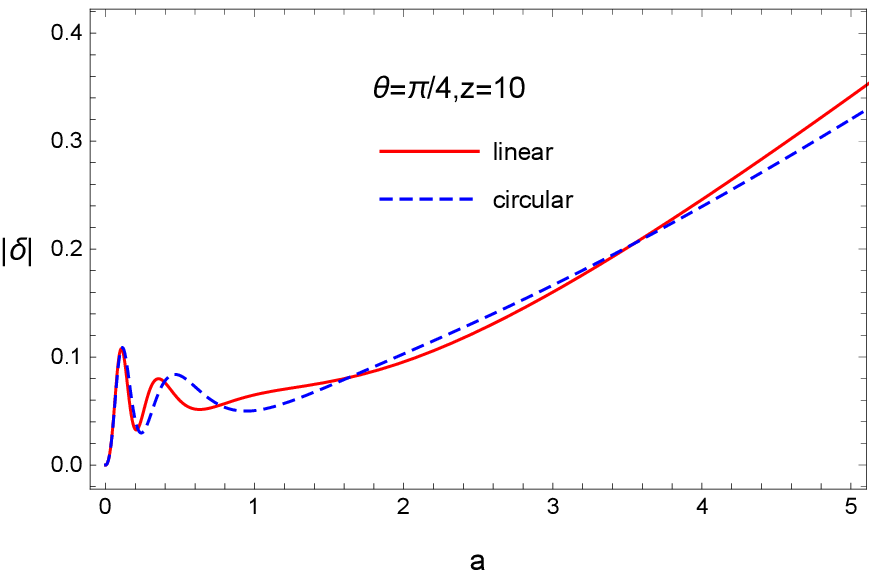}\vspace{0.0cm}
\includegraphics[scale=0.57]{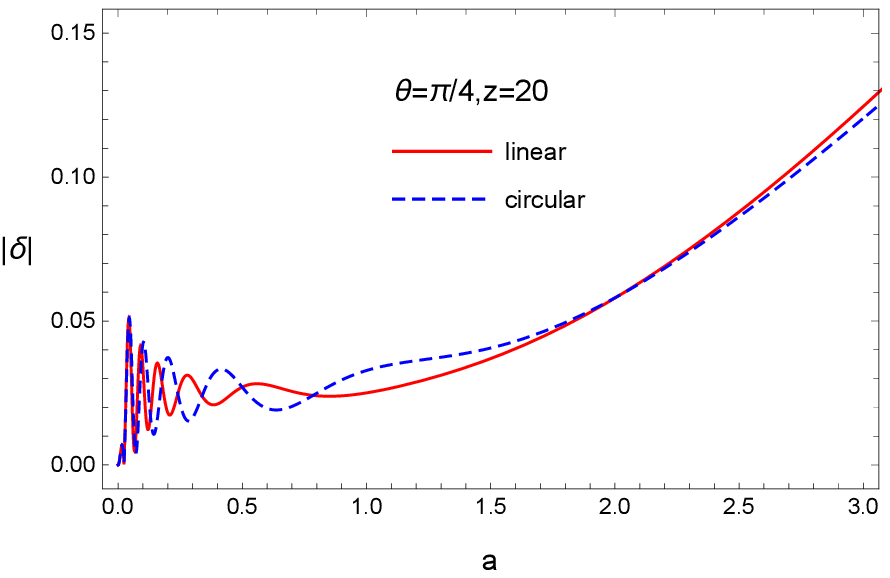}\\ \vspace{0.0cm}
\caption{\label{bgpa} Geometric phase as a function of acceleration for the initial atomic state $\pi/4$ with $z=0.5$, $z=10$, and $z=20$. The solid (red) and dashed (blue) lines represent the geometric phases purely due to linear acceleration case and circular acceleration case, respectively.}
\end{figure}

In Fig. \ref{bgpa}, we present the geometric phase as a function of acceleration for the initial atomic state $\theta=\pi/4$ with different $z$. Compared to the absence of a boundary, a larger value of geometric phase can be obtained. The geometric phase purely due to the linear acceleration case is larger than circular acceleration case for a large enough acceleration, although they fluctuate for small acceleration case.

\begin{figure}[ht]
\includegraphics[scale=0.57]{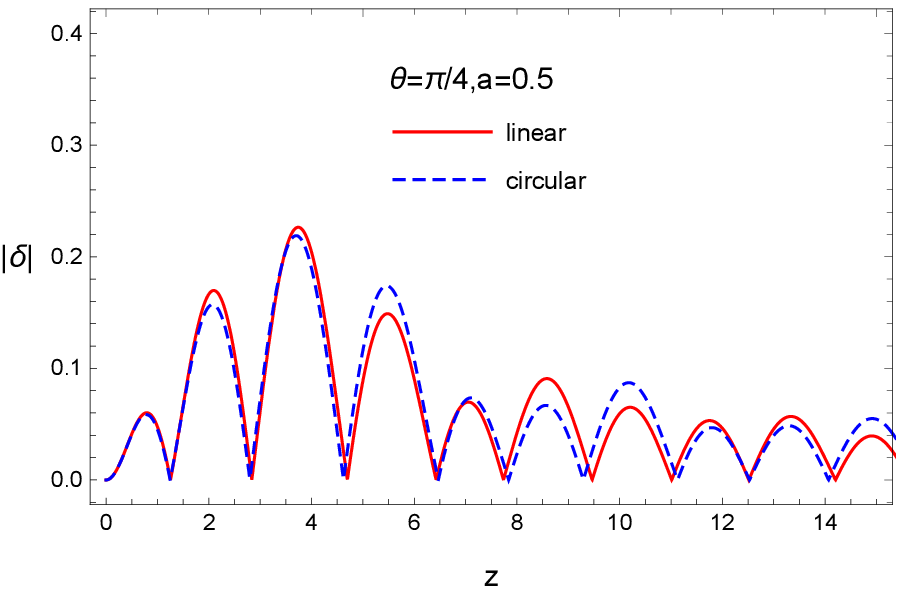} \vspace{0.0cm}
\includegraphics[scale=0.57]{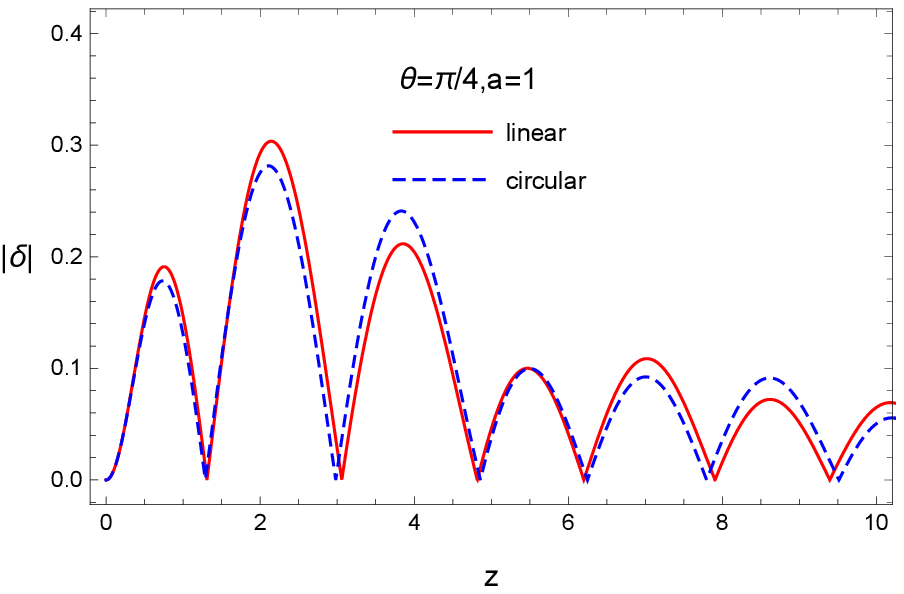}\vspace{0.0cm}
\includegraphics[scale=0.57]{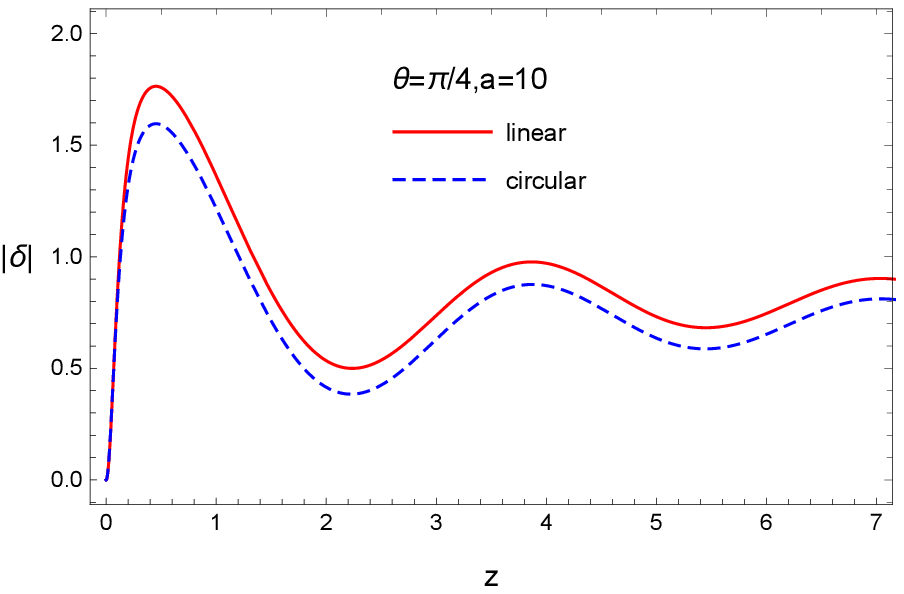}\\ \vspace{0.0cm}
\caption{\label{bgpz} Geometric phase as a function of $z$ for the initial atomic state $\theta=\pi/4$ with different $a$. The solid (red) and dashed (blue) lines represent the geometric phases purely due to linear acceleration case and circular acceleration case, respectively.}
\end{figure}

In Fig. \ref{bgpz}, we plot the geometric phase as a function of $z$ for the initial atomic state $\theta=\pi/4$ with $a=0.5$, $a=1$, and $a=10$. We find that the geometric phase fluctuates along $z$, and the maximum of geometric phase is closer to the boundary for a larger acceleration.

\begin{figure}[ht]
\includegraphics[scale=0.57]{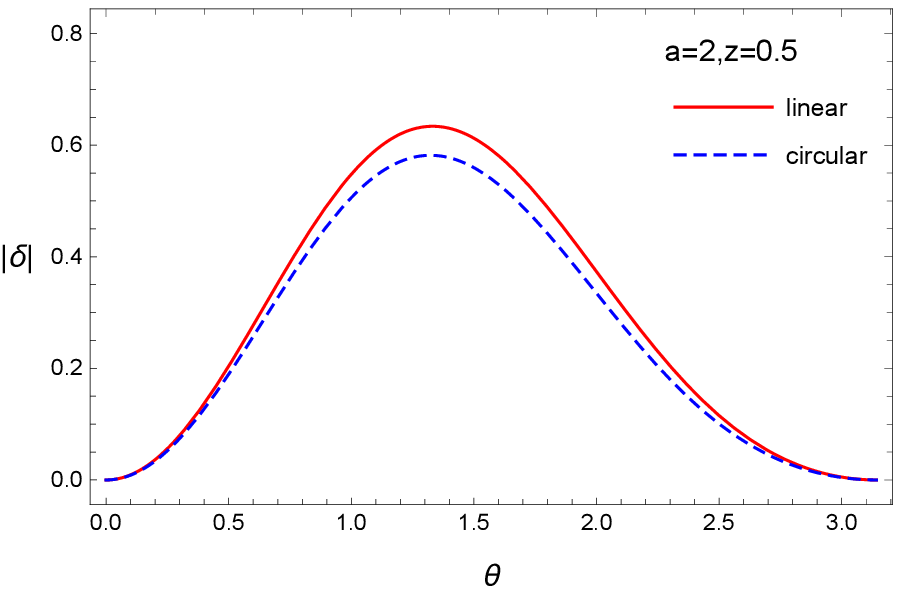} \vspace{0.0cm}
\includegraphics[scale=0.57]{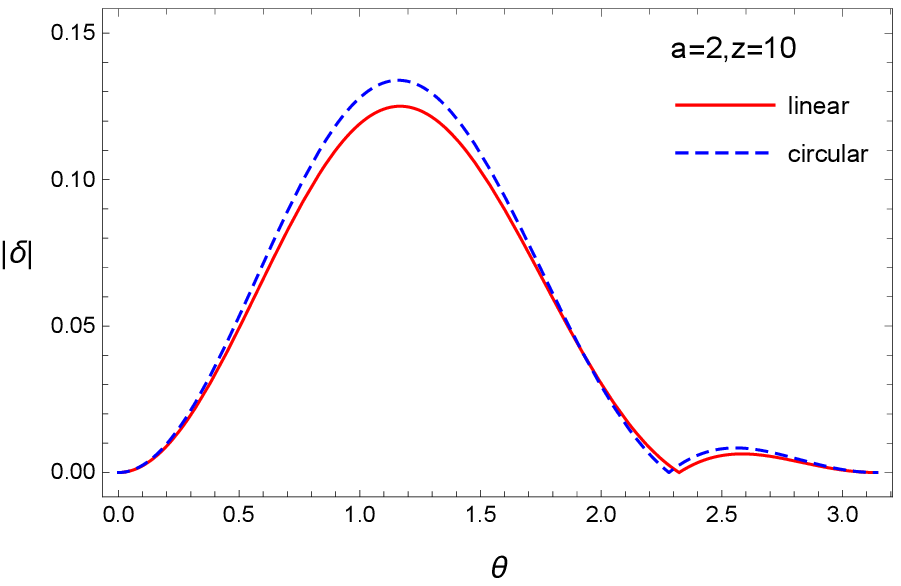}\vspace{0.0cm}
\includegraphics[scale=0.57]{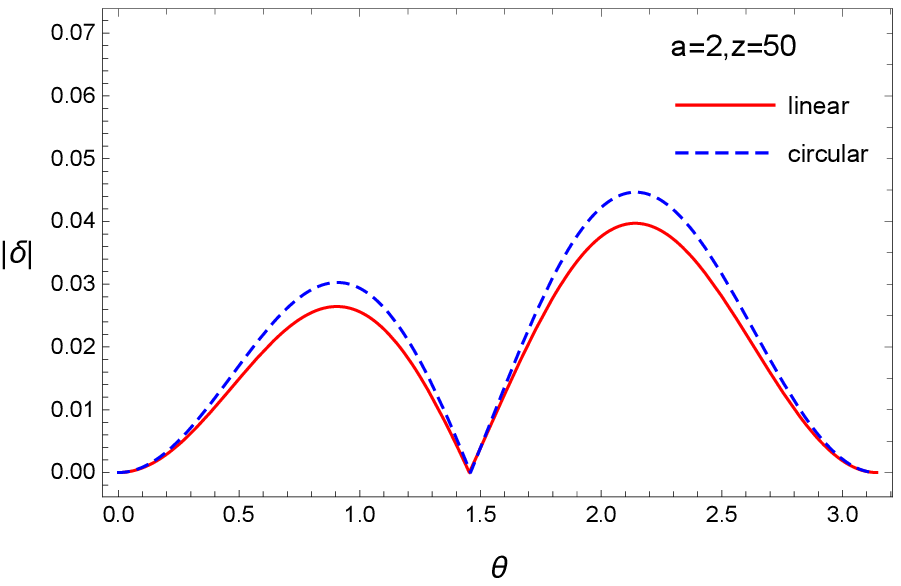}\\ \vspace{0.0cm}
\caption{\label{bgptheta} Geometric phase as a function of the initial atomic state with $a=2$ for $z=0.5$, $z=10$, and $z=50$. The solid (red) and dashed (blue) lines represent the geometric phases purely due to linear acceleration case and circular acceleration case respectively.}
\end{figure}

In Fig. \ref{bgptheta}, we depict the geometric phase as a function of the initial atomic state with $a=2$ for different $z$. We find that the geometric phase may be nonzero for $\theta=\pi/2$, which is different from unbounded case. The result implies that geometric phases can be acquired purely due to linear acceleration case and circular acceleration case with the initial atomic state $\theta \in(0,\pi)$ for a smaller $z$.

\section{Conclusions}

For a two-level atom interacted with a bath of fluctuating massless scalar fields in the Minkowski vacuum, we have investigated the geometric phases due to circularly accelerated case and linear acceleration. We found that the geometric phase acquired purely due to linear acceleration firstly is smaller than circular acceleration case in the ultrarelativistic limit for $\theta\in(0,\frac{\pi}{2})\cup(\frac{\pi}{2},\pi)$, then equals to the circular acceleration case in a certain $a$, and finally, is larger than the circular acceleration case. The spontaneous transition rates show similar feature. This result is different from the case of a bath of fluctuating vacuum electromagnetic fields that has been studied in Ref. \cite{Jin}. We concluded that the relation between linear acceleration case and circular acceleration case is not the same for different fluctuating vacuum fields. We also observed that the geometric phase acquired purely due to linear acceleration always equals to the circular acceleration case for the certain acceleration with every initial atomic state $\theta \in(0,\pi)$. One can use a circularly accelerated atom to simulate a linearly accelerated atom in a certain condition. With a boundary, we observed that a larger value of geometric phase can be obtained when compared to the absence of a boundary. We found that the geometric phase fluctuates along $z$, and the maximum of geometric phase is closer to the boundary for a larger acceleration. The geometric phases purely due to acceleration may be not zero for $\theta=\pi/2$, which is different from the unbounded case. The result suggested that, besides $\theta=\pi/2$, the geometric phases can be acquired purely due to linear acceleration case and circular acceleration case with the initial atomic state $\theta \in(0,\pi)$ for a smaller $z$.

\begin{acknowledgments}

This work was supported by the National Natural Science Foundation of China under Grant No. 11705144.

\end{acknowledgments}

\end{document}